\documentclass[%
 reprint,
superscriptaddress,
nofootinbib,
 amsmath,amssymb,
 aps,
]{revtex4-2}

\usepackage[utf8]{inputenc}
\usepackage{amsmath}
\usepackage{graphicx}
\usepackage{dcolumn}
\usepackage{bm}
\usepackage{siunitx}
\usepackage{cancel}
\usepackage[normalem]{ulem}
\usepackage{footnote}
\usepackage{float}
\usepackage{xcolor}
\usepackage{hyperref}

\begin{document}

\title{An interpenetrating-network theory of cytoplasm}

\author{Haiqian Yang}\email{hqyang@mit.edu}
\affiliation{Department of Mechanical Engineering, Massachusetts Institute of Technology, Cambridge, MA 02139, USA}
\author{Thomas Henzel}
\affiliation{Department of Civil and Environmental Engineering, Massachusetts Institute of Technology, Cambridge, MA 02139, USA}
\author{Eric M. Stewart}
\affiliation{Department of Mechanical Engineering, Massachusetts Institute of Technology, Cambridge, MA 02139, USA}
\author{Lallit Anand}
\affiliation{Department of Mechanical Engineering, Massachusetts Institute of Technology, Cambridge, MA 02139, USA}
\author{Ming Guo}
\affiliation{Department of Mechanical Engineering, Massachusetts Institute of Technology, Cambridge, MA 02139, USA}

\date{\today}
\begin{abstract}
Under many physiological and pathological conditions such as division and migration, cells undergo dramatic deformations, under which their mechanical integrity is supported by cytoskeletal networks (i.e. intermediate filaments, F-actin, and microtubules). Recent observations of cytoplasmic microstructure indicate interpenetration among different cytoskeletal networks, and micromechanical experiments have shown evidence of complex characteristics in  the mechanical response of the interpenetrating cytoplasmic networks of living cells, including viscoelastic, nonlinear stiffening, microdamage, and healing characteristics. However, a theoretical framework describing such a response is missing, and thus it is not clear how different cytoskeletal networks with distinct mechanical properties come together to build the overall complex mechanical features of cytoplasm. In this work, we address this gap by developing a finite-deformation continuum-mechanical theory with a multi-branch visco-hyperelastic constitutive relation coupled with phase-field damage and healing. The proposed interpenetrating-network model elucidates the coupling among interpenetrating cytoskeletal components, and the roles of  finite elasticity, viscoelastic relaxation, damage, and healing in the  experimentally-observed  mechanical response of interpenetrating-network eukaryotic cytoplasm.
\end{abstract}

\newcommand{\mat}{\text{\tiny R}}
\newcommand{\snd}{{\mathsf{d}}}
\newcommand{\trans}{{\mskip-2mu\scriptscriptstyle\top}} 
\newcommand{\tendot}{\mskip-3mu:\mskip-2mu}
\def\Gneq{G_{neq}}
\def\X{\mathbf{X}}
\def\x{\mathbf{x}}
\def\F{\mathbf{F}}
\def\FT{\mathbf{F^\trans}}
\def\A{\mathbf{A}}
\def\B{\mathbf{B}}
\def\C{\mathbf{C}}
\def\T{\mathbf{T}}
\def\P{\mathbf{P}}
\def\PT{\mathbf{P^\trans}}
\def\TT{\mathbf{T^\trans}}
\def\Tneq{\mathbf{T_{neq}}}
\def\Psieq{\Psi^{eq}_\mat(\bar{\C})}
\def\Psivol{\Psi^{vol}_\mat(J)}
\def\Psineq{\Psi^{neq}_\mat(\bar{\C},\{\A^{(i)}\})}
\def\Psineqi{\Psi^{neq(i)}_\mat(\bar{\C},\A^{(i)})}
\def\traction{\mathbf{t}}
\def\dv{{\tilde{\snd}}}
\def\chiv{\boldsymbol{\tilde{\chi}}}
\def\Fv{\mathbf{\tilde{F}}}
\def\xivec{\boldsymbol{\xi}}
\def\chivec{\boldsymbol{\chi}}
\def\n{\mathbf{n}}
\def\t{\mathbf{t}}

\maketitle
In nature, eukaryotic cells undergo large deformations in many essential physiological and pathological processes, such as cell division and migration taking place during morphogenesis, wound healing and cancer invasion~\cite{friedl2009collective,hu2019high,latorre2018Superelasticity,trepat2007universal,han2020cell}. When cells migrate through confined space, the mechanical deformations can be severe enough that lead to damage to nuclei and DNA, and even cell death~\cite{nuclearEnvelopeRupture,dnaDamage,nuclearEnvelopeRupture2,nuclearRupture}. Moreover, lung cells undergo large cyclic loading due to breathing~\cite{alcaraz2003microrheology}; skin cells are frequently exposed to mechanical injuries, while muscle cells constantly undergo cyclic deformation, which can lead to damage at a cellular level~\cite{abreu2012cytoskeleton}.

\begin{figure}[h]
    \centering
    \includegraphics[width=0.48\textwidth]{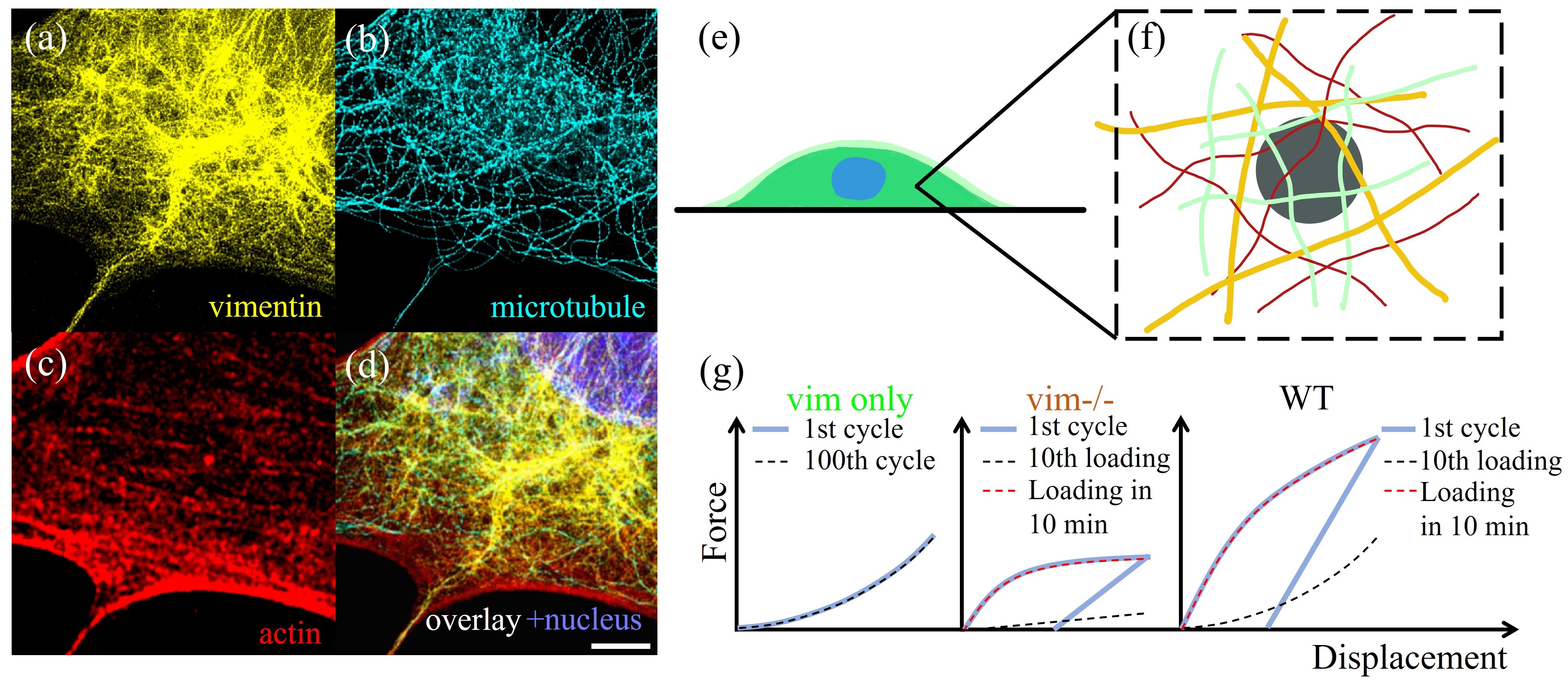}
    \caption{Cytoplasm is an interpenetrating network. Confocal image of (a) vimentin intermediate filament, (b) microtubule, (c) F-actin, and (d) overlay of a mouse embryonic fibroblast~\cite{hu2019high}. (Scale bar, \SI{5}{\micro\meter}). Schematics: (e) A cell. (f) The cytoplasm is supported by interpenetrating cytoskeletal fibers, which are intermediate filaments (green), F-actin (red), and microtubule (yellow). In a typical optical-tweezers measurement, a micro-size particle (grey) is perturbed within the cytoplasm. (g) Typical cyclic force-displacement response of vim only, vim-/-, and WT. Multiple loading-unloading cycles are applied by a laser-trapped particle. After 10 cycles, the particle stays in its original position for 10 minutes before another loading is applied.}
    \label{fig:schematic}
\end{figure}

The mechanical integrity and flexibility of the cytoplasm are supported by cytoskeletal networks~\cite{interpenetratingPRL,wu2022vimentin,wang1993mechanotransduction,janmey1991viscoelastic,hu2019high,wagner2007vimentin,ackbarow2007superelasticityVimentin,alisafaei2019regulation,zhang2020cytoskeletonModel,hang2021hierarchical,hang2022frequency,gupta2017equilibrium,fabry2001scaling,friedrich2012cells,fabry2003time,humphrey2002active,chaudhuri2007reversible,buxbaum1987shearThinning,broedersz2014modeling,gardel2004scaling,pegoraro2017mechanical,gardel2006prestressed,lin2011control,wang2001mechanical,ingber1997tensegrity}, which are interpenetrating networks formed by three major types of polymers: (i) intermediate filaments (vimentin, keratin and etc.), (ii) F-actin, and (iii) microtubules (Fig.~\ref{fig:schematic} (a-f))~\cite{interpenetratingPRL,wu2022vimentin}. The intermediate filament network is rather elastic, relatively tough and nonlinearly stiffens under external load, while F-actin and microtubule filaments show a relatively linear response before failure, and  they break quite easily \cite{janmey1991viscoelastic,hu2019high,wagner2007vimentin,ackbarow2007superelasticityVimentin}. Under large deformations, it is observed that the intermediate filaments can nonlinearly stiffen, while the F-actin and microtubule can relax, break and reform \cite{hu2019high}. The energy dissipation provided by the reorganization of the cytoskeletal polymer networks might be essential for cell survival under extreme mechanical deformations. Experimental observations of cell behavior under large deformations are  quite varied: while the force-displacement response of some cells exhibits stiffening at large deformations~\cite{fernandez2006stiffen,kasza2009stiffen,fernandez2008Harden}, others exhibit softening~\cite{trepat2007universal,lan2018fluidization}, and it is not immediately clear how the bulk mechanical properties of the cytoplasm arise from the combination of individual cytoskeletal networks. 

To elucidate the contributions from individual cytoskeletal networks to the experimentally-observed bulk mechanical response of eukaryotic cytoplasm, we develop a finite-deformation continuum-mechanical theory of the interpenetrating networks which includes visco-hyperelasticity coupled with phase-field damage and healing. We specialize the general theory to construct a minimal model; this model can capture the essential aspects of stiffening, relaxation, damage, and healing of cytoplasm which have been widely observed in mechanical experiments on eukaryotic cells.

\emph{ Summary of experimental observations.}---Recent \textit{in situ} micromechanical experiments by optical tweezers using eukaryotic cells of the wild type (WT) with all three types of the cytoskeleton, vimentin knock-out (vim-/-), and the vimentin-only ghost cell (vim only) have revealed (Fig.~\ref{fig:schematic}(g))~\cite{hu2019high}: 

\begin{itemize}
\item vim-only cells nonlinearly stiffen and they are relatively elastic under cyclic loading (Fig.~\ref{fig:schematic}(g), left).
\item Both WT and vim-/- cells relax, and multiple-cycle loading dramatically damages their load-carrying capacity (Fig.~\ref{fig:schematic}(g), middle, right).
\item The reaction force of the WT cells is generally larger than the sum of vim only and vim-/-, indicating an interaction among cytoskeletal networks.
\item Remarkably, in 10 minutes after 10 loading-unloading cycles, cells with F-actin and microtubule heal and restore the loading-bearing capacity. 
\end{itemize}

\emph{Summary of the theory}---To model these key experimental observations, we derive the governing equations from the principle of virtual power and free-energy imbalance~\cite{mao2018theory,gurtin2010mechanics} and present a summary of the minimal version of the theory here. A detailed derivation of the full version of the theory is provided in {\color{blue}Appendix~\ref{appendix: derivation}}.

We identify a body B with the region of space it occupies in a fixed reference configuration, and denote by $\mathbf{X}$ an arbitrary material point of B. A motion of B to the deformed body $\mathbf{\mathcal{B}}$ is then a smooth one-to-one mapping $\mathbf{x} = \boldsymbol{\chi}(\mathbf{X}, t)$ with deformation gradient given by~\footnote{We use the standard notation of modern continuum mechanics~\citep{gurtin2010mechanics}. Specifically: $\nabla$ and Div denotes the gradient and divergence with respect to the material point $\textbf{X}$ in the reference configuration, and  $\Delta =\text{Div} \nabla$ denotes the referential Laplace operator; grad  div, and div\, grad denote these operators with respect to the point $\textbf{x}=\boldsymbol{\chi}(\textbf{X},t)$ in the  deformed body; a superposed dot denotes the material time-derivative. Throughout, we write $\F^{-1} = (\F){}^{-1}$, $\F^{-\trans}=(\F)^{-\trans}$, etc. We write $\text{tr} \A$
for the trace
of a tensor $\A$. Also, the inner product of tensors $\A$ and $\B$ is denoted by $\A \tendot \B$, and the magnitude of $\A$ by $|\A|=\sqrt{\A \tendot \A}$. } 
\begin{equation}
\F = \nabla \boldsymbol{\chi}.
\end{equation}
%

The force balance is
\begin{equation}\label{eq: ForceBalanceMainText}
\begin{aligned}
           \text{Div}\, \P &= \mathbf{0},
\end{aligned}
\end{equation}
where the Piola stress $\P$  is given by
\begin{equation}
\P = 2\F \frac{\partial\Psi_\mat}{\partial \C},
\end{equation}
where $\Psi_\mat$ is the free energy per unit reference volume and $\C=\F^\trans\F$ is the right Cauchy-Green tensor.
%

To account for the  combined effects of nonlinear elasticity, visco-relaxation, damage, and healing in the interpenetrating networks of vimentin intermediate filaments with F-actin and microtubules, we propose the following form of the total free energy
\begin{equation}\label{eq:combineEnergyMainText}
\begin{split}
      \Psi_\mat &= \underbrace{\Psieq}_{\text{vimentin}}
        +\underbrace{g(\snd)}_{\text{damage}} \, \underbrace{\Psineq}_{\text{F-actin/microtubule}}\\
        &\\
        & + \Psivol + \Psi_\mat^{nonlocal}(\nabla \snd),
\end{split}
\end{equation}
with $J = \text{det} \, \F$ the volumetric deformation, $\bar\C= J^{-2/3}\C$ the distortional deformation tensor, $\{\A^{(i)}\}$ ($i=1, 2$) two tensorial symmetric and positive definite internal variables that quantify a long-time and a short-time visco-relaxation, $\snd$ a positive scalar damage variable ($\snd=0$ intact; $\snd>0$ damaged), and $\nabla \snd$ the gradient of the damage field. The degradation function $g(\snd)$ progressively damages the energy-carrying capacity of the F-actin/microtubule networks as the damage variable $\snd$ increases, and in this work takes the form of an exponential decay function\footnote{The body of literature concerning phase-field modeling of damage in solids has largely focused on modeling fracture processes, for which a widely selected degradation function is \citep{mao2018theory}
\begin{equation}
g(\snd) = (1-\snd)^2, \quad  \text{with} \quad \snd \in [0,1].
\end{equation}
Typically, also, the function $g(\snd)$ degrades the entire free energy of the material; in such a context $\snd = 0$ indicates intact material while $\snd = 1$ indicates a total loss of load-carrying capacity (i.e. the material is ``fractured''). For cytoplasm, however, we do not expect a complete loss of load-carrying capacity in the secondary networks even in the case of extensive damage. We therefore choose to interpret $\snd$ as a measure of mechanically-induced micro-damage to the secondary networks (e.g. force-induced rupture of polymer crosslinks~\cite{lee2010shearthinning2} in the F-actin and microtubules) and do not require it to be bounded between $0$ and $1$. Here we choose the degradation function 
\begin{equation}
g(\snd) = e^{-\snd}, \quad \text{with} \quad \snd\in [0,\infty),
\end{equation}
so that the secondary network degrades rapidly ($g'(\snd)$ is large) when $\snd$ is close to 0, and slowly ($g'(\snd)$ is small) as $\snd \to \infty$. We note that in the results presented in this paper, $\snd$ takes on values which are on the order of unity.}
\begin{equation}\label{eq:degradationMainText}
    g(\snd) = e^{-\snd}.
\end{equation}

The equilibrium energy $\Psieq$ accounts for nonlinear stiffening of elastic vimentin intermediate filaments, and is given by \citep{fung1967elasticity} 
\begin{equation}
 \Psi_\mat^{eq} = \frac{G_{eq}}{2b}\mathrm{exp} \left(b(\bar{I_1}-3)\right), \\
\end{equation}
with $G_{eq}$ the equilibrium shear modulus $b$ a nonlinear stiffening coefficient, and $\bar{I_1}= \text{tr} \bar\C$.

The non-equilibrium energy $\Psineq$ accounts for the viscoelasticity of F-actin and microtubule, and is given by \cite{linder2011micromechanically}
\begin{equation}\label{eq:PhiNEQMainText}
\Psi_\mat^{neq} = \sum_{i=1}^2 \frac{1}{2} \Gneq^{(i)} \left( \left(\A^{(i)}:\bar{\C}-3\right)-\text{ln} \, \left(\mathrm{det}\A^{(i)}\right) \right), 
\end{equation}
where $\Gneq^{(i)}$ is the non-equilibrium shear modulus of the $i$th viscoelastic branch, and $\A^{(i)}$ evolves according to 
\begin{equation}\label{eq:EvolutionAMainText}
\dot{\A}^{(i)} = \frac{1}{\tau^{(i)}}(\bar{\C}^{-1}-\A^{(i)}), \quad\quad (i=1,2),
\end{equation}
where $\tau^{(i)}$ is the relaxation time scale of the $i$th viscoelastic branch.

The volumetric energy $\Psivol$ accounts for the compressibility of the whole material, and is given by
\begin{equation}
    \Psi_\mat^{vol} = \frac{1}{2}\kappa(J-1)^2,
\end{equation}
with $\kappa$ the bulk modulus. In this work, we will treat the material to be nearly incompressible. Additional numerical details are provided in the {\color{blue}Appendix~\ref{appendix:numerical implementation}}.

The damage gradient energy $\Psi_\mat^{nonlocal}(\nabla \snd)$ accounts for energy stored in the process zone of the damaged F-actin and microtubule filaments, and is given by \citep{mao2018theory}
\begin{equation}\label{eq:NonlocalEAMainText}
\Psi_\mat^{nonlocal} = \frac{1}{2} \psi^*\ell^2 |\nabla \snd|^2,
\end{equation}
 where $\psi^*$ is the energy density stored in the damage process zone and $\ell$ is the length scale associated with the damage process zone.

To account for the damage and healing process of F-actin and microtubule networks, we propose the following form of the evolution equation for the damage variable $\snd$
\begin{equation}\label{eq:EvolutiondMainText}
    \underbrace{\zeta\dot{\snd}}_{\text{dissipative}} 
    = \underbrace{ e^{-\snd} \Psineq + \psi^*\ell^2 \triangle \snd}_{\text{energetic}} 
    - \underbrace{ \frac{\zeta}{\tau_H}\snd}_{\text{heal}},
\end{equation}
where $\zeta$ is a material parameter controlling the rate of damage and  $\tau_H$ is the healing time scale. This evolution equation states that damage $\snd$ in the non-equilibrium branch increases due to mechanical deformation and gradient effects, and decreases due to healing.  
%

\begin{table*}
\caption{\label{tab:table1}%
Model parameters. 
}
\begin{ruledtabular}
\begin{tabular}{cccccccccccc}
         &\multicolumn{2}{c}{Hyperelasticity}&\multicolumn{4}{c}{Viscoelasticity}& \multicolumn{2}{c}{Damage} & healing
\\
Cell type&\textrm{$G_{eq} (\SI{}{Pa})$}&\textrm{$b$}&\textrm{$G_{neq}^{(1)} (\SI{}{Pa})$}&\textrm{$\tau^{(1)} (s)$}&
\textrm{$G_{neq}^{(2)} (\SI{}{Pa})$}&\textrm{$\tau^{(2)} (s)$}&
\textrm{$\zeta (\SI{}{Pa \cdot s})$}&\textrm{$\psi^*\ell^2 (\SI{}{pN})$}&\textrm{$\tau_H (\SI{}{s})$}
\\
\colrule
WT       & 0.8  & 50    & 3.0  & 4.0  & 3.0  & 0.1   & 0.0003    &10  & 200 \\
vim-/-   & 0    & -     & 3.0  & 4.0  & 3.0  & 0.1   & 0.0003    &10  & 200 \\
\end{tabular}
\end{ruledtabular}
\end{table*}

\emph{Numerical implementation.}---We numerically implemented our theory in the finite element program FEniCS \cite{fenics1,fenics2}. Additional details concerning the numerical implementation can be found in {\color{blue}Appendix~\ref{appendix:numerical implementation}}. In our simulations, a rigid spherical particle (diameter $a = \SI{1}{\micro\meter}$) within a cylindrical matrix (diameter and height both \SI{20}{\micro\meter}) is displaced in the axial direction to a maximum distance of \SI{0.8}{\micro\meter} at a constant speed $\SI{1}{\micro\meter/s}$, then unloaded at the same speed to the initial position. Multiple-cycle loading-unloading is applied. Our simulations predict the force $F$ on the particle as a function of displacement $u$, which we then directly compare against force-displacement data measured in physcial experiments conducted on eukaryotic cytoplasm using optical tweezers in~\citep{hu2019high}.

\begin{figure}[h]
    \centering
    \includegraphics[width=.48\textwidth]{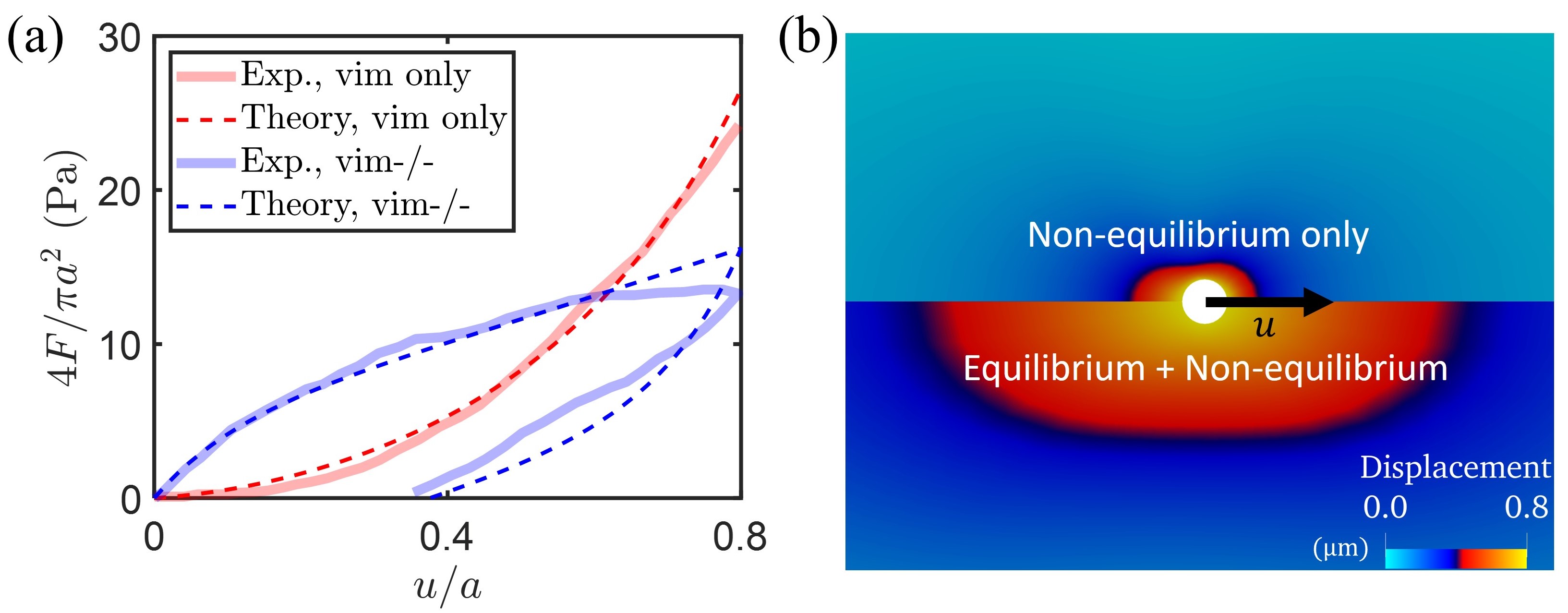}
    \caption{(a) The force-displacement relation in the vimentin-only cells is elastic, while the force-displacement relation in the vim-/- cells is dissipative. Experimental data from~\cite{hu2019high}. (b)The displacement-magnitude field at the maximum bead displacement ($u = \SI{0.8}{\micro\meter}$). The deformation is more diffuse in the interpenetrating network compared to a single non-equilibrium network. Nonlinear-stiffening network transduces long-ranged mechanical deformation inside the cytoplasm. The color map indicates the magnitude of displacement in the deformed configuration.}
    \label{fig:2}
\end{figure}

\emph{Individual network.}---Without any damage or healing, we first show that the micromechanical force-displacement response of vim-only ghost cells can be described by one single nonlinear equilibrium branch ($G_{eq}=\SI{0.3}{Pa}$ and $b=200$), while the force-displacement response of the first loading-unloading cycle of vim-/- cells can be reasonably well captured by two nonequilibrium branches, i.e. one slowly relaxing scale branch ($G_{neq}^{(1)}=\SI{1.2}{Pa}$ and $\tau^{(1)}=\SI{10}{s}$) and one fast-relaxing branch ($G_{neq}^{(2)}=\SI{5}{Pa}$ and $\tau^{(2)}=\SI{0.1}{s}$) (Fig. \ref{fig:2}(a)). {The minimal two-branch model is adequate for the cyclic loading-unloading at a constant speed considered in this work. It is worth noting that the weak power-law dependency of complex modulus on frequency~\cite{fabry2001scaling,gupta2017equilibrium,hang2021hierarchical,hang2022frequency} or the power-law relaxation~\cite{hu2019high} of cytoplasm can be approximated by a group of non-equilibrium branches.}

Interestingly, with the hyperelastic equilibrium branch added to the viscoelastic non-equilibrium branches, long-ranged deformation fields can be observed as was also observed in experiments, indicating more diffuse deformation through the nonlinear-stiffening vimentin network (Fig.~\ref{fig:2}(b)), consistent with experimental observations~\cite{hu2019high}.

\begin{figure}[h]
    \centering
    \includegraphics[width=.48\textwidth]{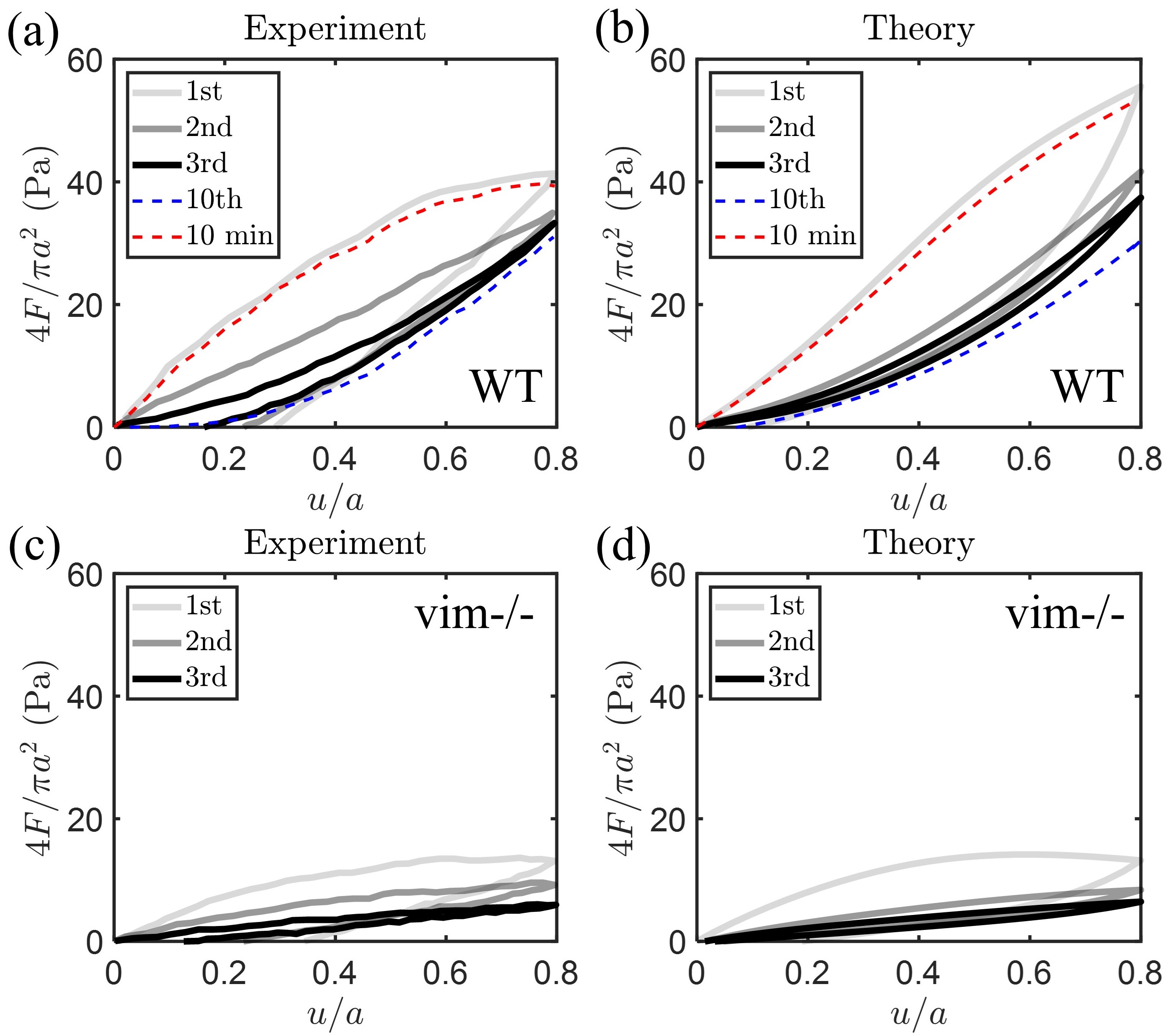}
    \caption{Damage and healing of interpenetrating-network cytoplasm in the WT cells the cytoplasm in the vimentin knock-out cells. Cyclic loading damages the viscoelastic network and reveals the elastic-stiffening network, in both (a) experiment and (b) theoretical prediction. The non-equilibrium network is damaged with 10 cycles of loading, while it almost fully heals in 10 minutes, in both (a) experiment and (b) theoretical prediction. 1st, 2nd, 3rd: the first, second and third loading-unloading cycle; 10th: the tenth loading; 10 min: loading in 10 minutes after the initial ten cycles are finished. Multiple-cycle damage of vim-/- cytoplasm in (c) experiment and (d) theoretical prediction. Experimental data from \cite{hu2019high}.}
    \label{fig:WTandKO}
\end{figure}

\emph{Wild-type interpenetrating cytoplasm.}---Moreover, we explore the damage and healing of WT cells with interpenetrating-network cytoplasm. With damage and healing enabled and the two-branch viscoelastic parameters around the damage-free model (Fig.~\ref{fig:2}(a)), we find that the model is capable of describing multiple-cycle hysteresis as well as healing of WT interpenetrating cytoplasm. The parameters are provided in TABLE~\ref{tab:table1}. We demonstrate that cyclic loading-unloading reduces the load-carrying capacity of the cytoplasm, and the force-displacement cycle gradually degrades from a viscous-dominated cycle to a hyperelastic-dominated cycle (Fig. \ref{fig:WTandKO}(b)), consistent with experiments shown in Fig. \ref{fig:WTandKO}(a). For the healing test, we hold the particle at the initial position for 10 minutes at the end of the 10-cycle loading-unloading, before another loading is applied. Remarkably, the cytoplasm restores its load-carrying capacity in 10 minutes, which is captured by the model (Fig. \ref{fig:WTandKO}(b)) and consistent with the experiment (Fig. \ref{fig:WTandKO}(a)).

\emph{Vimentin knock-out.}---We further test the predictive capabilities of the model by numerically ``knocking out" the vimentin intermediate filament network. To do so, we use exactly the same material parameters of WT and remove the hyperelastic branch (TABLE~\ref{tab:table1}). The numerical results nicely capture the behavior of the experimental results measured in vim-/- cells (Fig.~\ref{fig:WTandKO}(c\&d)). Note that consistent with experimental observations, the load-carrying capacity is dramatically reduced by removing the hyperelastic vimentin network (Fig.~\ref{fig:WTandKO}(c\&d)).

We note that in our numerical predictions, there is a region around the moving micro-bead with locally higher F-actin/microtubule damage ($\snd \approx 2$) and some non-zero damage throughout much of the computational domain since the secondary network damages easily. The spatial distribution of the damage depends upon several modeling factors, primarily: (i) the damage rate parameter $\zeta$, (ii) the nonlocal damage coefficient $\psi^*\ell^2$, and (iii) the presence of the nonlinear elastic vimentin network. However we do not have any experimental data with which to directly compare the distribution of damage predicted by the numerical implementation, as there is currently no experimental method for directly measuring the extent of damage to the secondary network.

\emph{Discussions.}--- Synthetic interpenetrating-network polymers composed of a tough background network such as polyacrylamide along with a brittle ``sacrificial" network such as alginate have previously been shown to possess mechanical toughness which is greater than the sum of each individual component~\cite{sun2012highly,gong2003double,gong2010double,haque2012super,ducrot2014toughening}. It is remarkable that eukaryotic cytoplasm seems to operate mechanically according to a similar principle, where the combination of a tough stretchable vimentin primary network and easily degradable actin/microtubule secondary network leads to enhanced toughness and stretchability~\cite{hu2019high}. 

Similar to the synthetic double-network polymers studied by \citeauthor{mao2017large} \citep{mao2017large}, we have shown that cytoplasm can indeed be modeled as interpenetrating networks. This interpretation of constituent networks helps to shed light on how different mechanical phenomena are built into the overall mechanical properties of the cytoplasm: intermediate filaments such as vimentin drive hyperelasticity and are responsible for the overall integrity of the material, while F-actin and microtubules are highly viscoelastic, and readily degrade to dissipate energy. Unlike the synthetic double-network polymers studied by \citep{mao2017large} however, the secondary network in living cells heals over time, which allows cells to better dissipate energy as they flexibly adapt their shape to the environment and then heal to recover their initial stiffness and toughness.

It has previously been suggested in the literature that the damage to the F-actin and microtubule networks (which we model with $\snd$) is related to the rupture of polymer crosslinks in these networks~\cite{lee2010shearthinning2,gupta2021optical}. It follows that healing processes in the cytoplasm might correspond to the reforming of these crosslinks, possibly mediated by ATP molecules. Such a damage-healing (unbinding-binding) process might be able to be visualized experimentally by fluorescent labeling of monomers and polymers of actin and microtubule. {Such experiments have been designed for synthetic interpenetrating networks where bond breaking can be visualized as damage propagates~\cite{ducrot2014toughening}. Such fluorescence experiments in the cytoplasm, if carried out, could yield detailed measurements of damage and healing which would be very useful for refining the specific forms of the degradation function and the healing function used in this work, both of which were taken to be of exponential character for simplicity.}

\emph{Concluding remarks.}---We have idealized the microstructure  of the cytoplasm in living cells as  an interpenetrating network of intermediate filaments, F-actin, and microtubules. Motivated by this physical picture, we then formulated a finite-deformation continuum-mechanical theory for the cytoplasm which comprises a tough, stretchable, and stiffening primary network in combination with a viscoelastic, damageable, and healable secondary network. We demonstrated that the theory and accompanying numerical implementation capture the micromechanical response, damage and healing of eukaryotic cytoplasm undergoing mechanical deformation via optical tweezers. These results help to interpret the interactions between the individual networks and explore the contributions from viscous relaxation, elastic stiffening, damage, and healing. The theory and accompanying numerical implementation represent significant advances in understanding and predicting the complex mechanical behavior of cytoplasm in living cells under large deformations. More generally, the theoretical framework and numerical implementation we have developed hold great potential for modeling cellular mechanical behaviors under large deformations involved in many other biological processes.

\bigskip
We would like to acknowledge the support from the NIH (1R01GM140108), the MathWorks, and the Jeptha H. and Emily V. Wade Award at the Massachusetts Institute of Technology. H.Y. acknowledges the MathWorks Mechanical Engineering Fellowship. M.G. acknowledges the Sloan Research Fellowship.
E.S. was supported by the Department of Defense (DoD) through the National Defense Science \& Engineering Graduate (NDSEG) Fellowship Program.

\newpage
\appendix
\section{Detailed derivation of the theory}\label{appendix: derivation}
\textit{Kinematics.} We identify a body B with the region of space it occupies in a fixed reference configuration, and denote by $\mathbf{X}$ an arbitrary material point of B. A motion of B to the deformed body $\mathbf{\mathcal{B}}$ is then a smooth one-to-one mapping $\mathbf{x} = \boldsymbol{\chi}(\mathbf{X}, t)$ with deformation gradient given by
\begin{equation}
\F = \nabla \boldsymbol{\chi}.
\end{equation}
We assume that
\begin{equation}
J = \text{det}\,\F>0,
\end{equation}
so that $\F$  is  invertible. 

The  symmetric and positive definite right Cauchy-Green tensor is defined as,
\begin{equation}
\C = \FT\F\,.
\label{cgree1}
\end{equation}
For later use we note that
 the deformation gradient $\F$ may be decomposed multiplicatively into volumetric and isochoric
factors $\F^v$ and $\bar\F$ as,
\begin{equation}
\F=\F^v\bar\F,
\end{equation}
where
\begin{equation}
\F^v=J^{1/3}\boldsymbol{1} \qquad\text{and}\qquad \bar\F=J^{-1/3}\F.
\end{equation}
This leads   to a multiplicative decomposition of the  
Cauchy-Green  tensor $\C=\FT\F$ of the form,
\begin{equation}
\C=\C^v\bar\C, \qquad \C^v=J^{2/3}\boldsymbol{1}, \qquad \bar\C=\bar\F^\trans\bar\F.
\end{equation}
We also  introduce  the invariant
\begin{equation}
\bar{I}_1 =  \text{tr}\, \bar\C.
\end{equation}

We further introduce a list of $N$ tensorial internal variables that quantify visco-relaxation\footnote{We have derived our theory in terms of an arbitrary number of viscous branches $N$, and for the results presented in the main body of this paper we select $N=2$.}
\begin{equation}
    \{\A^{(i)}\}, \quad (i=1, 2, ..., N),
\end{equation}
and a positive scalar damage phase-field ($\snd=0$ intact) 
\begin{equation}
    \snd, 
\end{equation}
and the gradient of the damage phase-field\footnote{Following recent phase-field damage literature, we also consider the gradient of the damage variable $\nabla\snd$ as a kinematical quantity \citep{mao2018theory}.}
\begin{equation}
   \nabla \snd.
\end{equation}

Throughout we denote by P an \emph{arbitrary} part of the reference body B, with $\n_\mat$ the outward unit normal on the boundary $\partial$P of P.

\textit{Virtual power.} We derive macroscopic and microscopic force balances via the principle of virtual power~\cite{mao2018theory,gurtin2010mechanics}. Consider a virtual motion (virtual configuration $\chiv$, virtual deformation gradient $\tilde{\F} = \nabla\chiv$, virtual damage $\dv$, and virtual damage gradient $\nabla \dv$) from an equilibrium state of the system. The principle of virtual power requires that the \emph{virtual} expenditure of internal power $W_{int}$ over an arbitrary part of the body P should be equal to the \emph{virtual} expenditure of external power  $W_{ext}$ over its boundary $\partial$P for any kinematically admissible virtual motion, viz.
\begin{equation}\label{eq:virtualPowerBalance}
    W_{ext}=W_{int}.
\end{equation}
Following \citep{mao2018theory}, the \emph{virtual} expenditures of internal and external power are assumed to be
\begin{equation}\label{eq:virtualPower}
\begin{aligned}
W_{int}&=\int_{\mathrm{P}} \left( \P:\Fv + \omega \dv+\xivec \cdot \nabla \dv \right)\,dv_\mat, \quad \text{and} \\
W_{ext}&=\int_{\partial \mathrm{P}} \left(\traction_\mat(\n_\mat) \cdot \chiv + \xi(\n_\mat) \dv \right) \,da_\mat,
\end{aligned}
\end{equation}
where $\P$ is the first Piola stress tensor which is power-conjugate to $\F$,  $\omega$ is the scalar micro-stress which is power-conjugate to $\snd$, and $\xivec$ is the vector micro-stress which is power-conjugate to $\nabla \snd$. Also, $\traction_\mat$ is the macroscopic surface traction in the reference configuration and  $\xi(\n_\mat)$ is the surface micro-traction associated with $\snd$.

Let $\dv=0$, then Eq.~\ref{eq:virtualPowerBalance} can be localized to find
\begin{equation}\label{eq:Force}
\begin{aligned}
    &\traction_\mat(\n_\mat) = \P \n_\mat \quad \text{and} \\
    &\text{Div} \, (\P) = 0.
\end{aligned}
\end{equation}

Consider an arbitrary virtual rigid rotation for any part P, given by 
\begin{equation} 
\Fv = \Omega \F \quad \text{and} \quad \dv = 0,
\end{equation}
where $\Omega$ is an arbitrary spatially constant skew tensor. The rigid motion hypothesis requires that $W_{int} = 0$, leading to~\cite{mao2018theory}
\begin{equation}
    \P \FT = \F\PT.
\end{equation}

Let $\Fv=0$, then Eq.~\ref{eq:virtualPowerBalance} can be localized to
\begin{equation}\label{eq:microforceBalance}
\begin{aligned}
    &\xi(\n_\mat) = \xivec \cdot \n_\mat, \quad \text{and} \\
    &\omega - \text{Div} \, (\xivec) =0. \\
\end{aligned}
\end{equation}

\textit{Free-energy imbalance.} Under isothermal conditions, the first two laws of thermodynamics reduce to the requirement that
\begin{equation}\label{eq:free energy imbalance}
    \int_{\mathrm{P}} \dot{\Psi}_\mat \, dv_\mat \leq W_{ext}.
\end{equation}

By using \eqref{eq:Force} and \eqref{eq:microforceBalance} and applying divergence theorem, we can write the \emph{actual} expenditure of external power $W_{ext}$ as
\begin{equation}\label{eq:divergenceTheorem}
\begin{aligned}
W_{ext} = \int_{\mathrm{P}} \left( \P:\dot{\F} + \omega \dot{\snd} + \xivec \cdot \nabla \dot{\snd} \right)\, dv_\mat.
\end{aligned}
\end{equation}
Using Eq.~\ref{eq:divergenceTheorem}, Eq.~\ref{eq:free energy imbalance} can then be localized as
\begin{equation}\label{eq:dissInequality}
    \dot{\Psi}_\mat-\P:\dot{\F}-\omega\dot{\snd}-\xivec \cdot \nabla \dot{\snd} \leq 0.
\end{equation}

We consider a free energy per unit reference volume $\Psi_\mat$ which depends on the list of constitutive variables $\boldsymbol{\Lambda}$ according to
\begin{equation}
    \Psi_\mat=\hat{\Psi}_\mat(\boldsymbol{\Lambda}) \quad
    \text{with} \quad 
    \boldsymbol{\Lambda} = \left\{ \F,\{\A^{(i)}\},\snd,\nabla \snd \right\}.
\end{equation}
%


\bigskip
\textit{Specialization.} We treat the cytoplasm material as interpenetrating networks with the two networks exhibiting the following major characteristics: 
\begin{itemize}
\item The primary vimentin intermediate filament network is ``tough'' in the sense that it remains intact under any deformation. We assume this network is purely elastic and nonlinearly stiffens at large stretch values. 
\item The secondary network, composed of interpenetrating F-actin and microtubules, is highly viscoelastic. This secondary network is easily damaged by mechanical deformation and exhibits healing over time. 
\end{itemize}

To account for the combined effects of nonlinear elasticity, visco-relaxation, damage, and healing in the interpenetrating networks of vimentin intermediate filaments with F-actin and microtubules, we propose the following form of the total free energy
\begin{equation}\label{eq: appEnergy}
\begin{aligned}
  &\hat\Psi_\mat(\boldsymbol{\Lambda})=\Psieq+g(\snd)\Psineq\\
  &\quad \quad \quad \quad +\Psivol +\Psi_\mat^{nonlocal}(\nabla \snd), \\
  & \quad \mathrm{with}  \quad \Psineq=\sum_{i=1}^N \Psineqi,
\end{aligned}
\end{equation}
where the equilibrium energy $\Psieq$ accounts for energy stored in the hyperelastic intermediate-filament network, the non-equilibrium free energy $\Psineqi$ accounts for the energy stored in the $i$th viscoelastic branch of the F-actin and microtubule network ($i=1,2,...N$), the volumetric free energy $\Psivol$ accounts for the slight compressibility of the whole material, and the damage gradient energy $\Psi_\mat^{nonlocal}(\nabla \snd)$ accounts for energy stored in the process zone of the damaged F-actin and microtubule filaments.

In writing \eqref{eq: appEnergy} we have introduced the degradation function $g(\snd)$, which damages the energy-carrying capacity of the secondary network as the phase-field parameter $\snd$ increases ($g(\snd) = 1$ intact, $g(\snd) = 0$ fully damaged). In this study, we shall choose
\begin{equation}
g(\snd)=e^{-\snd},
\end{equation}
which is monotonically decreasing with $\snd$.

Given the stiffening nature of the vimentin network, we assume the equilibrium free energy to be the Fung-type exponential energy function \cite{fung1967elasticity} in terms of distortional (i.e., isochoric) deformation
\begin{equation}
    \Psieq=\frac{G_{eq}}{2b}\mathrm{exp}(b(\bar{I_1}-3)),
\end{equation}
where $G_{eq}$ is an equilibrium shear modulus and $b$ is a stiffening coefficient.

We follow Linder et al.~\cite{linder2011micromechanically} and assume the free energy of each non-equilibrium branch takes the form
\begin{equation}\label{eq:noneqEnergy}
    \Psineqi=\frac{1}{2}\Gneq^{(i)} \left( (\A^{(i)}:\bar{\C}-3)-\text{ln} \, (\mathrm{det}\A^{(i)})\right),
\end{equation}
where $\Gneq^{(i)}$ is the non-equilibrium shear modulus of the $i$th branch.
The evolution equation for $\A^{(i)}$ is assumed to be~\cite{linder2011micromechanically}
\begin{equation}\label{eq:EvolutionA}
    \dot{\A}^{(i)} = \frac{1}{\tau^{(i)}}(\bar\C^{-1}-\A^{(i)}),
\end{equation}
where $\tau^{(i)}$ is the relaxation time scale of the $i$th viscoelastic branch.


We assume the volumetric energy to be quadratic
\begin{equation}
    \Psi_\mat^{vol} = \frac{1}{2}\kappa(J-1)^2,
\end{equation}
with $\kappa$ the bulk modulus.

We assume the nonlocal damage gradient energy to be quadratic \cite{mao2018theory}
\begin{equation}
    \Psi_\mat^{nonlocal}(\nabla \snd)=\frac{1}{2} \psi^*\ell^2 |\nabla \snd|^2,
\end{equation}
where $\psi^*$ is the energy density stored in the damage process zone, and $\ell$ is the length scale of the damage process zone.\footnote{For a damage process zone to be on the order of particle size ($\ell$ to be on the order of \SI{}{\micro\meter}), we expect $\psi^*$ to be on the order of the shear modulus Pa, such that the combined parameter $\psi^*\ell^2$ is on the order of pN.}

\textit{Constitutive relations.} We derive thermodynamically-consistent constitutive relations using the Coleman-Noll procedure~\cite{gurtin2010mechanics}. Using the chain rule, the free-energy imbalance Eq.~\ref{eq:dissInequality} can be written as
\begin{equation}\label{eq:fei2}
\begin{split}
    &\left(\frac{\partial \hat\Psi_\mat(\boldsymbol{\Lambda})}{\partial \F}-\P\right):\dot{\F}
     + \left(\frac{\partial \hat\Psi_\mat(\boldsymbol{\Lambda})}{\partial \snd}-\omega\right)\dot{\snd} \\
     & \quad + \left(\frac{\partial \hat\Psi_\mat(\boldsymbol{\Lambda})}{\partial \nabla \snd}-\xivec \right) \cdot \nabla \dot{\snd} + \sum_{i=1}^N \frac{\partial \Psineqi}{\partial \A}:\dot{\A} \leq 0,
\end{split}
\end{equation}
which must hold for all generalized motions $\{\dot{\F}, \dot{\snd}, \nabla{\dot{\snd}}, {\dot\A^{(i)}}\}$. To ensure satisfaction of \eqref{eq:fei2}, we shall first assume that $\P$ and $\xivec$ are given by
\begin{equation}\label{eq:microforce1}
\begin{aligned}
    \P &=  2\F \frac{\partial\hat\Psi_\mat(\boldsymbol{\Lambda})}{\partial \C}, \quad \text{and} \\
    \xivec &=\frac{\partial \hat\Psi_\mat(\boldsymbol{\Lambda})}{\partial \nabla \snd}, 
\end{aligned}
\end{equation}
and are thus both \emph{conservative} in the sense that their associated terms in \eqref{eq:fei2} are identically zero.

In contrast, to account for the dissipation and healing of damage in \eqref{eq:fei2}, we decompose $\omega$ into a conservative part and a non-conservative part according to 
\begin{equation}\label{eq:microforce2}
    \omega = \omega_{c} + \omega_{nc},
\end{equation}
where 
\begin{equation}\label{eq:omegaC}
\omega_{c} = \frac{\partial \hat\Psi_\mat(\boldsymbol{\Lambda})}{\partial \snd}.
\end{equation}
We further decompose the non-conservative part $\omega_{nc}$ into a dissipative part and a healing part according to
\begin{equation}\label{eq:ncDecomp}
\omega_{nc} = \underbrace{\zeta\dot{\snd}}_{\text{dissipative}} + \underbrace{\hat{\alpha}(\snd)}_{\text{healing}}, 
\end{equation}
where $\zeta$ is a rate-of-damage coefficient, and $\hat{\alpha}(\snd)$ is a positive-valued healing function to be specified later.\footnote{We could also include a constant on the right-hand side of 
\eqref{eq:ncDecomp} to account for a finite energy barrier which must be exceeded in order for damage to initiate. However, since we have no experimental evidence of an energy threshold for damage processes in the secondary network of cytoplasm, such a term is omitted in the current study.}

\textit{Damage evolution.} By combining the microforce balance Eq. \ref{eq:microforceBalance} and the constitutive relations Eq. \ref{eq:microforce1} and Eq. \ref{eq:microforce2}, we find that the evolution equation for the damage phase-field is
\begin{equation}\label{eq:Evolutiond}
    \zeta\dot{\snd} = e^{-\snd} \Psineq - \hat{\alpha}(\snd) + \psi^*\ell^2 \triangle \snd.
\end{equation}

\textit{Healing function.} In this work, for simplicity we assume that the healing process in the secondary network occurs as an exponential decay of the damage $\snd(t)$ over a characteristic healing time scale $\tau_H$.~\footnote{This might also be viewed as one spectrum of a possibly more complicated healing process by taking one term of the Laplace transformation at the given time of interest $\tau_H$.}

To this end, we specify a form of the healing function $\hat{\alpha}(\snd)$ which yields an exponential decay in damage $\snd(t)$ in the case of homogeneous damage with no mechanical deformation, where $\Psineq = 0$ and $\nabla \snd = \boldsymbol{0}$. In such a case, \eqref{eq:Evolutiond} reduces to the ordinary differential equation:
\begin{equation}\label{eq:ode}
    \zeta\dot{\snd} = - \hat{\alpha}(\snd).
\end{equation}
We use the healing function
 \begin{equation}\label{eq:healFunc}
 \hat{\alpha}(\snd)=\frac{\zeta}{\tau_H}\snd,
 \end{equation}
for which the solution to \eqref{eq:ode} is
\begin{equation}
\snd(t) \sim \text{exp} \, (-t/\tau_H).
\end{equation}
Finally, using \eqref{eq:healFunc}, the evolution equation for $\snd$ is 
\begin{equation}
    \zeta\dot{\snd} = e^{-\snd} \Psineq - \frac{\zeta}{\tau_H}\snd + \psi^*\ell^2 \triangle \snd.
\end{equation}

\textit{Discussion on dissipation and healing.} Using \eqref{eq:microforce1} and \eqref{eq:omegaC} in \eqref{eq:fei2}, we have the thermodynamic requirement 
\begin{equation}\label{eq:fei3}
-\omega_{nc} \, \dot{\snd}  + \sum_{i=1}^N \frac{\partial \Psineqi}{\partial \A}:\dot{\A} \leq 0.
\end{equation}
For the evolution equation \eqref{eq:EvolutionA}, it can be shown that the second term in \eqref{eq:fei3} is always smaller than zero~\cite{linder2011micromechanically}.
%
%
Similarly, because $\omega_{nc}$ is always greater than 0, in a damaging process $\dot{\snd}>0$ and therefore we have $\omega_{nc}\, \dot{\snd}>0$, so that \eqref{eq:fei3} is always satisfied. 

For a healing process ($\dot{\snd}<0$) however $\omega_{nc}\, \dot{\snd}$ is negative; it seems that in our theory \eqref{eq:fei3} may be violated for healing processes. However, the process of healing in the cell requires forming new F-actin/microtubule fibers by associating monomers with energy from ATP which we have not explicitly accounted for. It seems reasonable to expect that \eqref{eq:fei3} may still be satisfied by adding additional terms which account in some way for the energy input from ATP that drives the healing process. However, since we have no experimental basis for the specific form which these terms should take, in this work we omit them.

\textit{Governing equations. }
From Eq. \ref{eq:Force}, \ref{eq:EvolutionA}, and \ref{eq:Evolutiond} we have the governing equations
\begin{equation}\label{strongForms}
\begin{aligned}
    &\text{Div}\,(\P) = 0, \\
    &\zeta\dot{\snd} = e^{-\snd} \Psineq - \frac{\zeta}{\tau_H}\snd + \psi^*\ell^2 \triangle \snd,
\end{aligned}
\end{equation}
and $N$ evolution equations
\begin{equation}\label{evolution}
\dot{\A}^{(i)} = \frac{1}{\tau^{(i)}}(\bar\C^{-1}-\A^{(i)}),\quad\quad i=1,2,...,N,
\end{equation}
with each $\A^{(i)}$ initially set to the identity tensor.

This set of equations, supplemented with suitable boundary and initial conditions, can be solved for arbitrary boundary value problems.

\section{Some details of the numerical implementation}\label{appendix:numerical implementation}

We numerically implemented our theory in the finite-element program FEniCS \cite{fenics1,fenics2} using 2-D axisymmetric triangular elements. 

We solve the referential form of the governing equations \eqref{strongForms} as a combined weak (global) form using the finite element method. This combined weak form of the governing equations can be written as
\begin{equation}
     L_{mech}+L_{damage}=0,
\end{equation}
where $ L_{mech}$ is a weak form which accounts for mechanical governing equations, and $L_{damage}$ is a weak form which accounts for the damage-evolution governing equation.

In  our numerical implementation, we account for near incompressibility of the material by following a classical $(\mathbf{u},p)$ approach
 \citep[cf., e.g.,][]{bathe1996,  bonet1997} in which both the displacement $\mathbf{u}$ and a pressure-like field $p$ are used as degrees of freedom, with $p$ satisfying an additional governing equation~\footnote{We take $\kappa$ to be $10^3$~\SI{}{Pa} throughout the study, much larger than the typical shear modulus (\SI{\sim 1}{Pa}) considered.}
 \begin{equation}\label{eq:pStrongForm}
 p = \kappa \, (J-1).
 \end{equation}

Then, to specify the forms of  $ L_{mech}$ and $L_{damage}$, we first introduce a set of \emph{test functions} which correspond to each of the degrees of freedom as 
\begin{equation}\label{eq:testFxns}
\left\{ \mathbf{u}_{test}, \ p_{test}, \ \snd_{test} \right\}.
\end{equation}
The weak form of each governing equation is obtained by first multiplying the strong form by the corresponding test function from \eqref{eq:testFxns}, then integrating over the reference body B, and finally using the divergence theorem to reduce the order of derivatives if possible. Applying this process to the mechanical governing equations \eqref{strongForms}$_1$ and \eqref{eq:pStrongForm} yields the mechanical weak form~\footnote{We evaluate the derivative of the free energy in \eqref{eq:mechWeak} using the \texttt{diff()} function in FEniCS.}
\begin{equation}\label{eq:mechWeak}
    L_{mech} = \int_\text{B} \left( 2\F\frac{\partial \Psi_\mat}{\partial \C} : \nabla \mathbf{u_{test}} + \left(J-1-\frac{p}{\kappa}\right){p_{test}} \right) \, dv_\mat.
\end{equation}
By following a similar procedure and using a first-order finite difference scheme for the damage rate, in which
\begin{equation}
\dot{\snd} = \frac{\snd - \snd_{old}}{dt},
\end{equation}
the weak form of the damage phase-field evolution equation \eqref{strongForms}$_2$ is
\begin{equation}\label{eq:damWeak}
\begin{aligned}
    &L_{damage} = \int_\text{B} \Bigg( \Big[ \frac{\zeta}{dt}(\snd-\snd_{old}) - e^{-\snd} \Psineq \\
    & \qquad \qquad \qquad \qquad  + \frac{\zeta}{\tau_H}\snd\Big] \snd_{test} + \psi^*\ell^2 \nabla \snd \cdot \nabla \snd_{test} \Bigg) \, dv_\mat,
\end{aligned}
\end{equation}
where the variables with subscript ``$_{old}$'' are variables of the previous time step.

The evolution equation \eqref{eq:EvolutionA} for the internal tensor variables $\A^{(i)}$  is used to update $\A^{(i)}$ at each time step according to a backward Euler time integration scheme, viz.
\begin{equation}
\A^{(i)} = \left(1 + \frac{dt}{\tau^{(i)}} \right)^{-1} \, \left( \A^{(i)}_{old} + \frac{dt}{\tau^{(i)}} \bar\C^{-1} \right).
\end{equation}
%


%
Then, we damage the non-equilibrium free energy of the secondary network using the damage at the previous time step $\snd_{old}$ rather than $\snd$. That is, the numerical implementation has
\begin{equation}
  \Psi_\mat= \Psieq
        + \exp{(-\snd_{old})} \, \Psineq
        + \Psi_\mat^{nonlocal}(\nabla \snd),
\end{equation}
%
%

An illustration of the computational domain and simulation setup is shown in Fig.~\ref{fig:simBCs}. A 1-\SI{}{\micro\meter}-diameter particle is embedded in a large cylindrical matrix with both diameter and height \SI{20}{\micro\meter}. The lateral surface of the matrix is fixed in the axial direction. The rest of the boundaries are traction-free. Without being explicitly stated, a zero-valued Neumann boundary condition is naturally assigned to the damage field.~\footnote{In fact, in writing the weak forms \eqref{eq:mechWeak} and \eqref{eq:damWeak}, we have omitted any boundary terms since these terms are all zero in the simulations presented in this paper.} In our simulations, the particle is rigidly displaced from its initial position and the resultant force is measured.

A GitHub repository which contains example code which will generate the results reported for the ``wild-type" cell in Figure 3(b) of this paper is available online: 
\begin{itemize}
\item \url{https://github.com/ericstewart36/cytoplasm}
\end{itemize}


\renewcommand\thefigure{\thesection \arabic{figure}}
\setcounter{figure}{0}
\begin{figure}
    \centering
    \includegraphics[width=.3\textwidth]{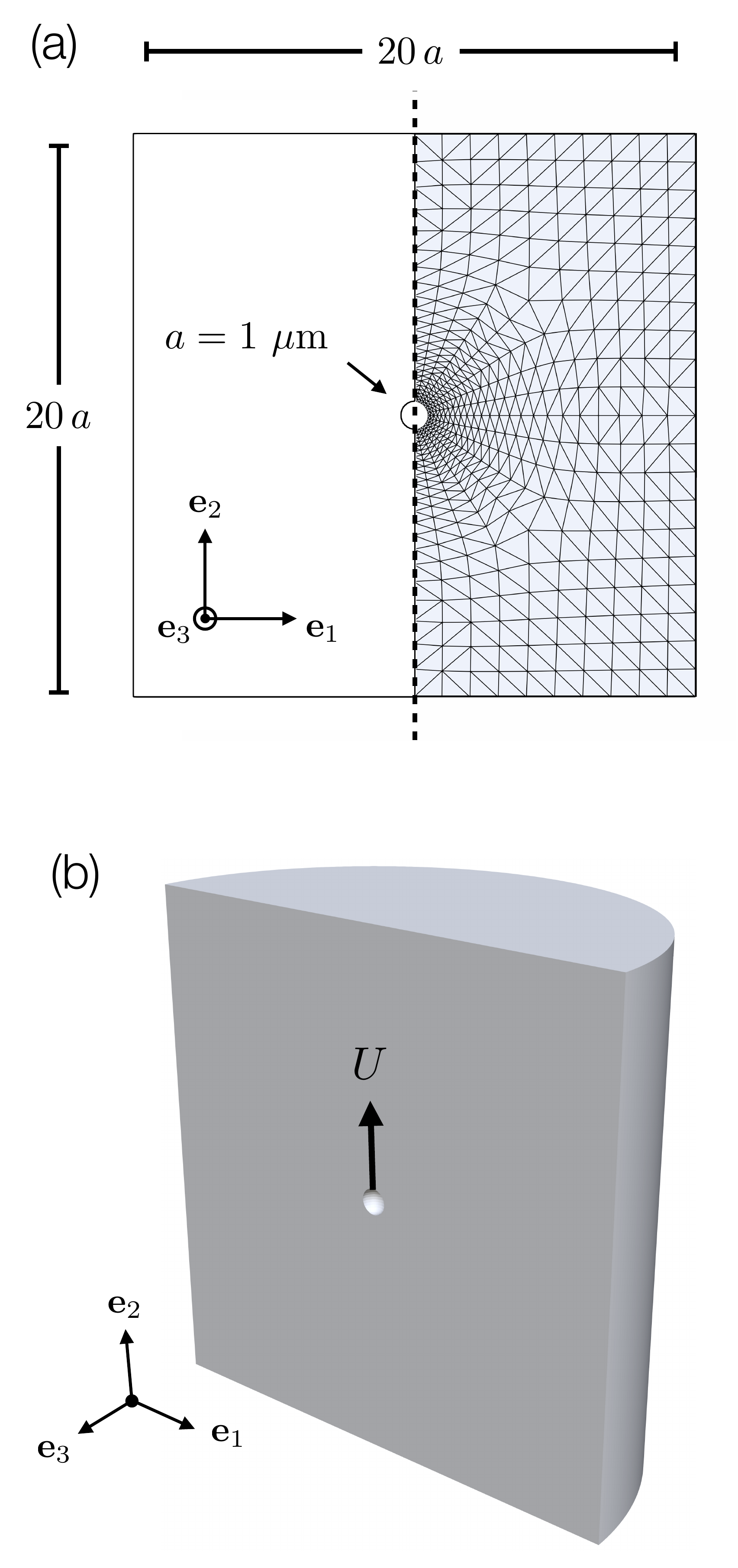}
    \caption{(a) The computational domain and mesh used in our simulations, where the dotted line indicates the axis of rotational symmetry. (b) A cutaway view of the full 3-D domain, formed by a 180-degree rotation of the mesh about the axis of symmetry. A rigid displacement $U$ is applied to the micro-particle in the $\boldsymbol{e}_2$-direction, and the resulting reaction force is measured.}
    \label{fig:simBCs}
\end{figure}

\newpage
\bibliography{reference}

\end{document}